\documentclass[runningheads]{llncs}

\usepackage{dialogue}
\usepackage{tabularx}
\usepackage{float}
\usepackage{amsmath}
\usepackage{subcaption} 
\usepackage[T1]{fontenc}
\usepackage{graphicx}
\usepackage{multirow}
\usepackage{enumitem}
\usepackage{xcolor}
\usepackage{orcidlink}

\usepackage{tabularx} %
\usepackage{booktabs} %
\usepackage{makecell} %
\usepackage{tabularx}
\usepackage{multirow}
\usepackage{booktabs}
\usepackage{makecell}
\usepackage{float} 
\usepackage{multirow}
\begin{document}
\title{Modernizing Ground Truth: Four Shifts Toward Improving Reliability and Validity in AI in Education}
\titlerunning{Four Shifts Toward Improving Reliability and Validity} %

\author{
Danielle R. Thomas\orcidlink{0000-0001-8196-3252}\inst{1} \and
Conrad Borchers\orcidlink{0000-0003-3437-8979}\inst{1} \and
Kirk P. Vanacore\orcidlink{0000-0003-0673-5721}\inst{2} \and
Kenneth R. Koedinger\orcidlink{0000-0002-5850-4768}\inst{1} \and
René F. Kizilcec\orcidlink{0000-0001-6283-5546}\inst{2}
}

\authorrunning{Thomas et al.}

\institute{
Carnegie Mellon University\\
\email{\{drthomas,cborchers,koedinger\}@cmu.edu}
\and
Cornell University\\
\email{\{kpv27,kizilcec\}@cornell.edu}
}

\maketitle 

\begin{abstract}

Generative Artificial Intelligence (GenAI) is now widespread in education, yet the efficacy of GenAI systems remains constrained by the quality and interpretation of the labeled data used to train and evaluate them. Studies commonly report inter-rater reliability (IRR), often summarized by a single coefficient such as Cohen's kappa ($\kappa$), as a gatekeeper to ``ground truth.'' We argue that many educational assessment and practice support settings include challenges, such as high-inference constructs, skewed label distributions, and temporally segmented multimodal data, which yield potential misapplication or misinterpretation of threshold-based heuristics for IRR. The growing use of large language models as annotators and judges introduces risks such as automation bias and circular validation. We propose four practical shifts for establishing ground truth: (1) treat IRR as a diagnostic signal to localize disagreement and refine constructs rather than a mechanical acceptance threshold (e.g., $\kappa$ $>$ 0.8); (2) require transparent reporting of rater expertise, codebook development, reconciliation procedures, and segmentation rules; (3) mitigate risks in LLM annotation through bias audits and verification workflows; and (4) complement agreement statistics with validity and effectiveness evidence for the intended use, including uncertainty-aware labeling (e.g., assigning different labels to the same item to capture nuance), criterion-related checks (e.g., predictive tests to check if labels forecast the intended outcome), and close-the-loop evaluations of whether systems trained on these labels improve learning beyond a reasonable control. We illustrate these shifts through case studies of multimodal tutoring data and provide actionable recommendations toward strengthening the evidence base of labeled AIED datasets.

\end{abstract}

\section{Introduction}

Humans often make notoriously flawed judgments. Research separates these human errors into two categories: \textit{noise} and \textit{bias} \cite{hilbert2012toward,kahneman2021noise}. \textit{Noise} is the random variability in judgments of the same or highly similar case \cite{kahneman2021noise}, whereas \textit{bias} is systematic directional error. While social bias and algorithmic fairness have received considerable attention \cite{baker2022algorithmic,kizilcec2022algorithmic}, this work argues that \textit{noise} is an underexamined but consequential obstacle for the AI in Education (AIED) community. 

Education, at its best, is inherently noisy. Assigning grades, defining ``student engagement,'' and identifying ``giftedness'' are tasks laden with subjective interpretation. For instance, consider the task of labeling a tutor’s feedback: one expert educator might code a hint as “effective scaffolding,” while another, with an equally valid pedagogical perspective, could see it as “giving away the answer.” Yet, AIED often relies on inter-rater reliability (IRR) metrics, frequently summarized by a single coefficient such as Cohen’s $\kappa$, which are often applied and interpreted as if they provide an objective gatekeeping test of ``ground truth,'' even in settings involving high-inference constructs, skewed label distributions, or contextual judgment  \cite{gwet2014handbook,mchugh2012interrater,thomas2025beyond}. In such settings, disagreement does not necessarily indicate annotator error; it may reflect systematic differences in interpretation, underspecified constructs, or valid pedagogical ambiguity. Valid pedagogical ambiguity refers to the inherent complexity of educational interactions where multiple, equally defensible interpretations exist based on pedagogical perspectives or theoretical frameworks. Historically, the community has borrowed standards from machine learning, where disagreement is often treated as error to be minimized and high agreement is implicitly treated as a proxy for label quality \cite{ando2005framework}.

A troubling trend in recent AIED literature is the treatment of reliability as a perfunctory checkbox rather than a rigorous scientific process. Many studies fail to report internal reliability measures entirely. For example, a meta-systematic review of AI in higher education found that 51.5\% of the articles lacked any reliability metrics or coding protocols \cite{bond2024meta}. Even when included, reporting is often superficial, typically restricted to standalone agreement scores (e.g.,  \cite{carpenter2020detecting,wong2025rethinking,zhuang2025integrating}). Another problematic practice is ``resolution by consensus,'' where researchers state that disagreements were discussed until a specific $\kappa$ threshold was achieved (e.g., \cite{zhao2025autograding}). Without detailing how disagreements were resolved, this process can amount to forcing consensus on complex constructs, effectively deleting valid pedagogical ambiguity from the data and making it difficult to determine if a model is learning the construct or merely annotators' biases. Without explicit documentation of how disagreements map to construct definitions, high post-hoc agreement may provide little evidence that the resulting labels support valid interpretation, rather than simply reflecting shared annotator heuristics.

Reconsideration of how our field labels and annotates data is timely. The rapid proliferation of Generative AI has ushered in a new era for AIED, promising scalable AI-driven tutoring systems \cite{mittal2024comprehensive}. In particular, multimodal interaction data, such as human-AI tutoring systems and classroom discourse analysis, are of increasing interest \cite{aleven2023towards,thomas2025advancing,wong2025rethinking}. However, the learning efficacy of these systems depends on a fundamental bottleneck: the quality of human (and, recently, AI) annotations. If an AI model is trained on ``ground truth'' data labeled by inconsistent annotators, the resulting tool may simply codify and replicate that inconsistency \cite{messer2024consistent}. A model trained on noisy or weakly specified labels may achieve high accuracy in reproducing annotator judgments, while failing to support the instructional mechanisms required to improve learning. Beyond immediate instructional efficacy, such models may be limited in their ability to support theory or yield interpretable estimates aligned with learning constructs. This aligns with cautions from EDM that models prioritizing prediction accuracy often fall short in providing interpretable, meaningful estimates useful for theory building \cite{Rachatasumrit2024Beyond}. Similarly, long-held warnings within qualitative psychology contend that the significance of a specific pattern is not strictly defined by its quantifiable prevalence, but instead by its capacity to capture a fundamental and meaningful element of the phenomenon being studied \cite{braun2006using}.

This challenge intensifies with the controversial emergence of large language models (LLMs) as judges, co-raters, or data augmentors \cite{chen2024humans}. Efficient yet contentious (c.f., \cite{davison2024ethics,jowsey2025we}), LLMs complicate both the interpretation of agreement and the evidentiary role that agreement plays in validation, particularly when models act as annotators, judges, or data generators. We advocate replacing one-size-fits-all reliability metrics with methods that respect the inherent subjectivity of learning. To that end, we explain the unique landscape of AIED and contribute four shifts providing concrete actions to improve reliability practices: {\textbf{1){ Reframe IRR from a threshold to a diagnostic tool.}}} We argue for moving beyond mechanical acceptance thresholds (e.g., $\kappa>0.8$) to using reliability metrics as signals for identifying productive disagreement and valid pedagogical ambiguity; {\textbf{2) Contextualize and enforce transparent reporting.}} We outline requirements for reporting rater characteristics, codebook evolution, and conflict resolution, while addressing AIED challenges like the ``linking problem'' in multimodal data; \textbf{{3) Mitigate risks in LLM annotation.}} We propose auditing methods to detect ``beauty'' and ``authority'' biases and call for verification frameworks (e.g., simulatability) when using LLMs as synthetic judges. \textbf{{4) Prioritize validity over consensus.}} We argue for moving beyond one-size-fits-all interpretations of reliability metrics towards clearer separation between reliability evidence, validity evidence, and evidence of instructional effectiveness.

\section{Related Work}
To assess consistency in human judgment, researchers commonly report inter-rater reliability (IRR), which quantifies the extent to which independent raters produce similar labels under a given coding scheme \cite{cole2024inter}. Metrics like Cohen's kappa ($\kappa$) estimate agreement beyond chance under specific assumptions about label distributions and rater behavior \cite{mchugh2012interrater}. However, treating all disagreement as random noise is reductive; disagreement may reflect underspecified constructs, differences in pedagogical perspective, or contextual interpretation rather than unreliability. Disagreement often stems from systematic measurement error or valid ambiguity in high-inference and subjective tasks. Recent work argues that rather than eliminating this variance, researchers should treat ``disagreement as data,'' analyzing differences in how multiple raters explain their annotation reasoning to better understand complex educational constructs and using ambiguity to guide opportunities for refinement \cite{tajik2026disagreement}. 

Cohen’s $\kappa$ is a common metric that improves upon simple percent agreement by accounting for chance, providing a standardized scale for reliability comparisons \cite{mchugh2012interrater}. While traditional heuristics categorize $\kappa$ values in ranges from “poor” ($<$ 0.4), “substantial” (0.61-0.80), to “almost perfect” (0.81-1.0) \cite{landis1977application}, some educational studies adopt heuristics, treating $\kappa$ values below 0.60 as potentially insufficient depending on task complexity and intended use \cite{mchugh2012interrater}. However, the utility of a given $\kappa$ score is highly dependent on the downstream application of the annotated data. A low $\kappa$ does not automatically render a label useless. For applications that require stable, generalizable estimates across raters and contexts, persistently low agreement poses clear risks. By contrast, for applications such as curating illustrative cases for feedback or professional development, a ``low $\kappa$, useful subset'' scenario is possible. In such cases, the specific instances where raters do agree, even if infrequent, can represent highly useful data. Further, when raters from diverse pedagogical backgrounds reach a consensus on a difficult, high-inference construct, it likely signals an exceptionally clear and intelligible example of that construct. This highlights the importance of reporting rater characteristics, as such context is often necessary to interpret agreement estimates and assess whether consensus reflects construct clarity or shared perspective \cite{ocumpaugh2015baker}. Other common IRR measures include Fleiss’ $\kappa$, an adaptation of Cohen's $\kappa$ that can be used when there are more than two raters making judgments on a nominal scale; and Krippendorff’s $\alpha$, a highly flexible and robust IRR metric that can handle any number of raters, different types of data (nominal, ordinal, interval, ratio), and can accommodate missing data \cite{mchugh2012interrater}. These measures address different design constraints (e.g., number of raters, scale type), but none by themselves establish the validity of the underlying construct being annotated.

In AIED, IRR is often treated as a primary gatekeeping signal for labeled data quality. Low IRR can indicate substantial variability in judgment, though the source of that variability may include noise, construct ambiguity, or rater heterogeneity. High IRR alone can create a false sense of evidentiary sufficiency for validity \cite{gwet2014handbook,Rachatasumrit2024Beyond,thomas2025beyond}. A team of annotators can be perfectly reliable (high agreement) while sharing the same systematic bias or misconception (low validity). For example, limiting “student engagement” to “raising hands” introduces bias in the measurement step by distilling a rich state of the world into a narrow value \cite{kizilcec2022algorithmic}. While annotators may achieve high IRR, the AI will systematically miss quiet engagement, illustrating that reliability captures consistency of labeling, not whether the labels support valid interpretation of the intended construct.

\section{The Unique Landscape of AIED}
AIED data---ranging from student open-ended responses and behavioral clickstream data to biometric signals like eye-tracking and facial expressions---are fundamentally dual-purpose, differing from data in many other domains. It is used for both theory-building to uncover cognitive and social mechanisms of learning, similar to psychology \cite{self1990theoretical}, and to propel instructional action, such as providing real-time feedback in an intelligent tutoring system \cite{koedinger2012knowledge}. This dual responsibility necessitates a nuanced approach to measurement for both low- and high-inference tasks, presenting challenges that are amplified in AIED relative to many traditional machine learning benchmarks \cite{rosenshine1970evaluation,thomas2025beyond}.

\textbf{\textit{AIED contains a wide range of low- and high-inference tasks.}} \textit{Low-inference tasks} involve the tallying of direct, denotable behaviors that require minimal interpretation, such as ``Did the tutor mention the student’s name?'' or ``How many words did the student speak?'' In these cases, traditional agreement metrics like Cohen’s $\kappa$ are often high because the constructs are physically observable or easily inferrable. Conversely, educational constructs of interest, such as ``confusion'' (see Fig. 1), ``conversational uptake,'' or ``productive struggle,'' are often \textit{high-inference tasks} \cite{demszky2021measuring}, meaning they require contextual interpretation and grounding in higher-order concepts that are not directly observable from individual acts or behaviors. Unlike computer vision tasks where labeling an image as a ``cat'' is a low-inference perceptual task, labeling a student's ten-second pause as ``deep thinking'' versus ``distraction'' requires deep contextual and pedagogical interpretation. While a high-inference rating (e.g., a 1–5 scale of scaffolding quality) is often more strongly associated with downstream learning outcomes \cite{ocumpaugh2015baker}, it is significantly harder to reach a high IRR because humans bring different perspectives and experiences, which influence evaluation. Similarly, the context (e.g., cultural, social) of the data might make the annotation of high-inference tasks harder for raters. There are numerous high-inference tasks in AIED.

\textbf{\textit{Reliability values vary widely by use case.}} Historical recommendations for $\kappa$ range widely from 0.4 to 0.8 \cite{landis1977application}, often failing to account for the distribution of the data itself. In high-inference domains where ``ground truth'' is uncertain and coding is holistic, such as affect detection, lower $\kappa$ scores are expected and often acceptable. For instance, the Baker Rodrigo Ocumpaugh Monitoring Protocol (BROMP), a rigorous field observation method for quantifying student engagement, sets its certification threshold at $\kappa$ $>$ 0.6 \cite{ocumpaugh2015baker}. This reflects the reality that for subjective constructs like student affect, achieving ``perfect'' agreement is less critical than achieving sufficient aggregate reliability across a large volume of data. Conversely, for low-inference tasks with clear definitions, higher thresholds are both expected and necessary. This variance underscores that reliability is not an intrinsic property of the data alone, but depends on the construct definition, annotation design, and intended use. Table 1 illustrates examples of use cases, reported acceptable IRR, and the task inference level.

\begin{table}[htbp]
    \centering
    
    \captionsetup{font=footnotesize} %
    
    \caption{Use cases and applications containing reportedly acceptable IRR measures and inference level of task. Reported IRR thresholds vary widely across tasks.}
    \label{tab:irr-use-cases}
    
    \fontsize{6pt}{7.6pt}\selectfont %
    \renewcommand{\arraystretch}{0.95} 
    
    \begin{tabularx}{\textwidth}{ >{\raggedright\arraybackslash}X l >{\raggedright\arraybackslash}X }
        \toprule
        \textbf{Use case/application} & \textbf{Acceptable IRR} & \textbf{Inference level of task} \\
        \midrule
        Agreement between a trainee and certified trainer needed to allow the trainee to begin coding data \cite{ocumpaugh2015baker} & $\kappa = 0.6$ & Low to high inference. Behavior and affect coding schemes. Affective states are considered high-inference. \\
        \midrule
        Agreement between two human coders providing scores of learners open responses within online lessons on advocacy \cite{thomas2025does} & $\kappa = 0.55-0.93$ & Low to high inference. Assessing tutors’ ability to respond to various tutoring situations varies in subjectivity. \\
        \midrule
        Agreement between human experts and foundation LLMs (e.g., Gemini 3 Pro, GPT-5) on coding Talk Moves in tutoring dialogues \cite{vanacore2025large} & $\kappa = 0.38-0.58$ & High inference. LLMs struggled with tagging tutor moves that required sensitivity to social dynamics and metacognitive contexts. \\
        \midrule
        Agreement between multimodal LLMs and expert educators on grading handwritten student work \cite{henkel2025seeing} & \makecell[tl]{$\kappa = 0.90$ (Arithmetic)\\ $\kappa \approx 0.47$ (Illustrations)} & Low to high inference. Reliability drops from objective arithmetic (low) to interpreting student drawings (high). \\
        \bottomrule
    \end{tabularx}
\end{table}

\textit{\textbf{AIED often uses sequential and multimodal data.}} Educational data is frequently sequential and multimodal (e.g., classroom audio and video). Even when annotators agree that a specific pedagogical event occurred, they often disagree on its precise onset and offset times, a phenomenon known as the ``linking problem'' \cite{holle2015easydiag}. Standard reliability metrics often penalize these minor timestamp offsets heavily, reporting ``poor'' reliability even when human judgments are conceptually aligned. This sensitivity to segmentation differences can create an illusion of low data quality in otherwise valid datasets, specifically in tasks involving visual-audio discrepancies. In such cases, low reliability values may be deemed acceptable for specific analytic purposes, or may indicate metric failure rather than annotator error. For example, researchers found a significant discrepancy between high raw agreement scores (often exceeding 90\%) and low chance-corrected $\kappa$ scores for instructional activity labels in classroom videos \cite{foster2024automated}. The authors attribute this paradox to the timed-event nature of the data, where slight temporal misalignments deflate $\kappa$ scores despite conceptual consensus. This highlights why a rigid interpretation of standard measures, such as the widely cited guideline that $\kappa$ $<$ 0.60 is ``unacceptable'' \cite{mchugh2012interrater}, can be misleading in multimodal educational contexts. For complex, event-based data, relying solely on a single $\kappa$ value is often insufficient; researchers may need to consider a combination of metrics or adjust acceptability thresholds to account for the ``linking problem'' inherent to sequential analysis. Table 2 illustrates an example of the “linking problem” in real-life multimodal tutoring data.

\begin{table}[htbp] %
\centering
  \captionsetup{font=footnotesize} %
\caption{An example of temporal latency in multimodal tutoring. The lag between visual and audio events leads to misaligned annotations.}
\label{tab:temporal-latency}

\fontsize{6pt}{8pt}\selectfont %
\renewcommand{\arraystretch}{0.9} %

\begin{tabularx}{\textwidth}{|>{\raggedright\arraybackslash}X|c|c|}
\hline
\multicolumn{1}{|l|}{\textbf{Multimodal transcription}} & \multicolumn{2}{c|}{\textbf{Student error?}} \\
\cline{2-3}
& \textbf{Coder 1} & \textbf{Coder 2} \\
\hline
\textcolor{red}{[14:53]} [\textbf{visual event}]: `Sorry, incorrect...' feedback appears. The correct answer is -12g - 9.1 & X & X \\
\hline
\textcolor{red}{[14:55]} [\textbf{visual event}]: New problem loads: -4u +-6u + -2u -5u] &  & \\
\hline
\textcolor{red}{[14:55]}  \textbf{Tutor}: It's okay. Okay, what about this one? Can we combine or how many terms do we have or are in a relationship with somebody? & X &  \\
\hline
\textcolor{red}{[15:03]} \textbf{Student}: We have four...actually three, cuz there's three that are negative and one that is positive. & & X \\
\hline
\textcolor{red}{[15:08]} \textbf{Tutor}: Okay, Does that mean we can combine them or we can't combine them? & & \\
\hline
\textcolor{red}{[15:12]} \textbf{Student}: We can. & & \\
\hline
\textcolor{red}{[15:13]} \textbf{Tutor}: Yeah, combine them. Perfect. & & \\
\hline
\textcolor{red}{[15:15]} \textbf{Tutor}: Okay. So what do we get when we put those all together? & & \\
\hline
\textcolor{red}{[15:31]} \textbf{Student}: 17u. & X & \\
\hline
\textcolor{red}{[15:33]} [\textbf{visual event}]: `Sorry, incorrect...' appears. The correct answer is -17u. & X & X \\
\hline
\textcolor{red}{[15:36]} \textbf{Tutor}: Not quite. & & X \\
\hline
\end{tabularx}
\end{table}

\subsection{Case Study Context: Multimodal Tutoring Data}
Effective tutoring is among the most powerful influences on student learning \cite{nickow2020impressive}, yet understanding effective tutor moves has been stalled by a lack of granular data \cite{thomas2025advancing}. Decades of research have lacked the large-scale, multi-modal datasets of tutoring and teaching (video, audio, transcripts, assessments, and metadata) necessary to unpack these dynamic interactions at the level of in-the-moment tutoring moves. We focus on multi-modal tutoring because the field is now seeing a proliferation of rich interaction data (e.g., National Tutoring Observatory, SafeInsights, SEERNet) that is increasingly processed by LLMs and AI agents to automate interpretation. Multi-modal tutoring interaction is particularly well suited for examining reliability challenges because the ``ground truth'' is rarely contained in a single modality. Fig. 1 illustrates an example of multi-modal tutoring interaction data, which we use as a running example to ground our methodological arguments. The example shows how a student’s verbal confusion (“I don’t know what to do here”) is paired with visual cues (e.g., shrugging, pointing to an equation) and processed into a transcribed, timestamped interaction. AI models operationalize these signals by inferring latent states (e.g., confusion), resolving ambiguous references (e.g., “here” to $x^2$ = 225), and classifying tutor moves, such as prompting self-explanation, affective support, and cognitive scaffolding. If annotation quality is evaluated solely through expert agreement statistics, we risk validating models that primarily mimic human consistency, including our biases and blind spots, rather than capturing the latent processes that drive student learning. This case highlights why agreement statistics alone are insufficient for validating annotations intended to support explanatory research and instructional intervention.

\begin{figure}[ht]
    \centering
     \captionsetup{font=footnotesize} %
    \includegraphics[width=1\textwidth]{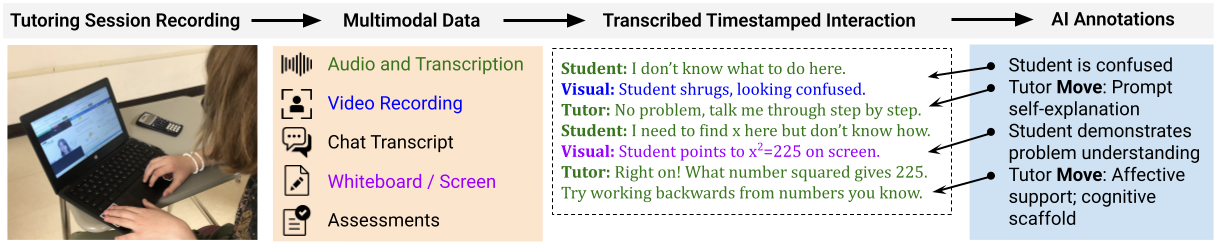}
    \caption{A processing pipeline for multimodal tutoring data. Raw audiovisual recordings are transformed into transcripts enriched with AI-extracted features, highlighting the complexity of labeling pedagogical moves.}
    \label{fig:line-graph}
\end{figure}

\section{Four Shifts Toward Improving Reliability}
\subsection{Reframe IRR from Mechanical Threshold to a Diagnostic Tool}
We propose reframing how IRR is used and interpreted in AIED: rather than treating a single coefficient as a threshold for acceptance (e.g., $\kappa$ $>$ 0.8), we emphasize diagnostic use that is sensitive to construct complexity, annotation design, and intended application. We aim to better align IRR uses with low-to-high inference educational constructs and the integration of LLMs. \cite{cole2024inter} provided a description of what IRR can and cannot do for qualitative case study research. We adapt this framing to AIED by highlighting common failure modes in multimodal and high-inference annotation and by linking agreement statistics to concrete design and reporting decisions. Table 3 summarizes common misinterpretations of IRR in AIED practice and contrasts them with defensible uses of agreement statistics as evidence about consistency, rather than as substitutes for validity or effectiveness claims.

\begin{table}[htbp]
\centering
\captionsetup{font=footnotesize} %
\caption{What inter-rater reliability (IRR) can and cannot do for AIED with examples from multimodal tutoring and educational discourse literature. Extended from \cite{cole2024inter} to reflect the nature of educational annotation and LLM integration.}
\label{tab:irr-can-cannot}

\fontsize{6pt}{7.6pt}\selectfont %
\renewcommand{\arraystretch}{1} %

\begin{tabularx}{\textwidth}{ 
  >{\raggedright\arraybackslash}X  %
  >{\raggedright\arraybackslash}X  %
  r                             %
}
\toprule
\textbf{IRR is...} & \textbf{IRR is not...} & \textbf{Ex.} \\
\midrule
\textbf{{A diagnostic tool}} to determine the degree of consensus between coders as a basis for discussion that may enhance the findings or theory emerging from the data.
&
\textbf{{A mechanical threshold}} (e.g., $\kappa > 0.8$) that must be cleared to prove a dataset is ``ready'' for machine learning.
&
\cite{vitale2025from} \\
\midrule

\textbf{{A potential signal of productive disagreement}}, where variance between human and/or LLM coders may warrant construct refinement or further validation.
&
\textbf{{A reason to eliminate disagreement}}, as it may reveal the need for construct refinement, additional rater training, or complementary validity evidence.
&
\cite{petukhova2025intent} \\
\midrule

\textbf{{A measure of consistency}} for low-inference tasks (e.g., ``Did the student ask a question?'') to high-inference tasks (e.g., ``Is the student engaged?'') to ensure data quality for baseline AI models.
&
\textbf{{A guarantee of validity.}} High agreement on a biased rubric (e.g., engagement = ``quietness'') only ensures the AI will accurately replicate that flaw.
&
\cite{ahtisham2025ai,ocumpaugh2015baker} \\
\midrule

\textbf{{A starting point}} for understanding human coders and LLM agreement, used to assess variation.
&
\textbf{{An ending point}} or sole indicator of AI accuracy. Matching a human coder does not mean the AI is ``right.''
&
\cite{vanacore2025large} \\
\midrule

\textbf{{A tool for transparency}} in the annotation pipeline, revealing how a codebook evolved through reconciliation to handle disagreements and edge cases.
&
\textbf{{A way to turn educational measurement into a mechanical task}} that ignores the context of the teacher/tutor-student relationship.
&
\cite{ahtisham2025ai,barany2024chatgpt} \\
\bottomrule
\end{tabularx}
\end{table}

\subsection{Contextualize and Enforce Transparent Reporting}
Reliability measures need to be reported and the process clearly described. A troubling trend in recent AIED literature is the treatment of reliability as a perfunctory checkbox rather than a rigorous scientific process. Many studies fail to report internal reliability measures entirely \cite{bond2024meta}. Even when included, reporting is often surface level, typically restricted to standalone $\kappa$ scores \cite{carpenter2020detecting,wong2025rethinking}, leading to the common trope that ``raters coded 20\% of the data and agreement was good ($\kappa$ = 0.8).'' This reliance on  simple percent accuracy masks the need for granular, chance-corrected measures necessary to accurately capture reliability (e.g., \cite{li2025asd,zhuang2025integrating}). A particularly problematic practice is the ``resolution by consensus'' description, where researchers state that ``disagreements were discussed until a $\kappa$ of 0.80 was achieved.'' Without detailing how these disagreements were resolved, this process often amounts to forcing consensus on high-inference constructs, effectively deleting valid pedagogical ambiguity from the dataset. This opacity prevents the field from understanding whether a model is learning the construct or merely the biases of the annotators. Similarly, when reviewing papers, reviewers should not automatically criticize work that reports lower than ideal reliability measures, if thorough explanation and justification is included.

AIED can benefit from greater reliance on domain expertise in construct definition, rater training, and adjudication—paired with transparent documentation of how expert judgments are elicited and reconciled. For many high-inference educational constructs, “ground truth” is unlikely to be adequately approximated by minimally trained crowd-workers; domain experts can improve construct specification and adjudication, especially when their roles and procedures are transparently documented. For instance, the Baker Rodrigo Ocumpaugh Monitoring Protocol (BROMP) enforces rigorous certification of human observers to ensure reliability before data collection even begins \cite{ocumpaugh2015baker}. We argue that AIED should normalize this level of rigor. This entails two shifts: 1) \textbf{{increased transparency}}, reporting not just the final $\kappa$, but the demographics of the raters, the specific codebook definitions, and the nature of the disagreements found; and 2) \textbf{{inclusion of experts}}, by enrolling experienced educators or domain-experts not merely as annotators, but as co-authors whose pedagogical expertise defines the constructs we aim to capture.

\subsection{Mitigate Risk in LLM Annotation via Auditing and Verification}
LLMs are shifting data annotation from a manual task to an efficient human-AI collaboration, recasting them as active agents—especially in AIED where domain expertise is scarce \cite{tan2024large}. We categorize their roles as an equal-party annotator, a judge, and a data augmentor.

\textbf{\textit{LLM as an Equal-Party Annotator}}. One recent use for LLMs is that of an equal-party annotator, where the LLM functions as a distinct rater alongside human raters. This approach is particularly attractive for scaling the coding of large educational datasets, such as classroom transcripts or MOOC discussion forum posts, where using purely human-coded data is cost-prohibitive and time consuming \cite{tan2024large,vanacore2025large,petukhova2025intent}. Research suggests that for context-independent and low-inference tasks (e.g., identifying explicit questions or social tags), LLMs can achieve high agreement with humans \cite{hou2024prompt,na2025llm}. However, research indicates that for context-dependent constructs (e.g., “theorizing” or “reflection”), base models often struggle to reach acceptable reliability without extensive fine-tuning or the use of the largest frontier models \cite{hou2024prompt,thakur2025judging}. Furthermore, traditional metrics like percent agreement often mask poor model performance. While LLMs can serve as a ``first pass'' coder to reduce cognitive load, they introduce risks of automation bias. \cite{wang2024human} demonstrated that providing LLM pre-labels can actually decrease human accuracy when the model is incorrect, as annotators may uncritically adopt the AI’s suggestion. Effective human and LLM collaboration may require a verification framework (e.g., the Lapras system), where a separate verifier model identifies low-confidence LLM labels for human re-annotation \cite{ahtisham2025ai,wang2024human}. 

\textbf{\textit{LLM as a Judge.}} In the ``LLM-as-a-judge''  framework, AI models evaluate natural language responses like student essays or tutoring dialogues, offering a scalable, low-cost alternative to human annotation that has been shown to approximate crowd-workers agreement in some constrained settings \cite{nahum2025llms}. This practice, however, introduces the critical risk of \textit{circular validation}, a process where one AI's output is validated by another AI, creating a self-referential loop that can provide a false sense of accuracy \cite{chen2024humans}. Instead of verifying genuine correctness, such a loop often serves only to amplify the models' shared biases and flawed reasoning. LLM judges are also susceptible to systemic biases, such as ``authority bias'' (overvaluing citations) and ``beauty bias'' (favoring technical formatting) \cite{chen2024humans}, or ``position bias'' in pairwise comparisons \cite{li2024llms}. For instance, in multimodal math, models excel at rote arithmetic ($\kappa$ = 0.90) but fail on conceptual illustrations ($\kappa \approx$  0.47), showing technical skill does not equal pedagogical ``tacit knowledge'' \cite{henkel2025humans}. Beyond these biases, LLMs also show 
distinct “grading personalities,'' meaning high human-LLM agreement might merely reflect the human adapting to the training data \cite{henkel2025humans}. For educational validity, an LLM's rationale must satisfy \textit{simulatability}, meaning a human can predict its judgment from the explanation \cite{wang2024human}. High agreement can mask flawed reasoning; even when LLMs achieve high IRR, their explanations often diverge from human reasoning in ambiguous cases \cite{henkel2025humans}. This ``right for the wrong reasons'' phenomenon shows the danger of using score alignment as a sole proxy for pedagogical validity.

To justify replacing human annotators despite these risks, researchers have proposed various statistical procedures, such as the ``Alternative Annotator Test'' \cite{calderon2025alternative}. This test verifies whether an LLM offers a better alternative to recruiting a human by comparing the model with a small group of annotators (at least three) on a modest subset of data (50–100 examples). By demonstrating that the LLM aligns with the group consensus as well as or better than an individual human, researchers can assess whether an AI model offers a cost-effective substitute under defined assumptions about quality, risk, and oversight. These validity concerns extend beyond annotation to the use of LLMs for data augmentation, where models generate “synthetic ground truth” to address data scarcity \cite{tan2024large}. While scalable, this practice creates a risk of “recursive training,” where models amplify surface-level stylistic biases while losing pedagogical precision \cite{thakur2025judging}. Consequently, LLM outputs should be evaluated using multiple complementary forms of evidence (targeted audits, agreement diagnostics, and predictive validity checks), to ensure alignment with educational mechanisms and outcomes.

\subsection{Complement Agreement with Validity and Impact Evidence}
To prioritize educational impact over statistical uniformity, we outline complementary sources of evidence that operate at three levels: representation, validation, and intervention. These evidence sources are not interchangeable; rather, they address distinct threats to validity and misuse at AIED pipeline stages.

\textbf{\textit{Multilabel Annotation.}} An annotation that allows for more than one label relaxes the assumption of a single, definitive ``correct'' label when constructs are ambiguous or multimodal evidence is incomplete. Instead, it allows for multiple valid interpretations by assigning several labels to a single data point. This method is particularly useful for capturing the complexity and ``productive ambiguity'' inherent in many educational tasks \cite{thomas2025beyond}. An example of applying this to multimodal tutoring data is shown in Table 4. Here, an annotator identifies student errors through multiple simultaneous valid lenses: \textit{explicitly}, via visual system logs (e.g., ``Sorry, incorrect'' appearing on screen) and \textit{implicitly}, via the tutor's verbal cues (e.g.,``not quite'' or ``it is okay''). The schema includes a confidence level, which allows the annotator to signal when the ground truth is ambiguous rather than forcing a low-confidence binary decision. This multilabel approach preserves the nuance of how errors manifest across different modalities, rather than relying on a single, potentially incomplete annotation.

\vspace{-6mm}

\begin{table}[] %
\centering
\captionsetup{font=footnotesize} %
\caption{A multilabel annotation example of student math errors. The annotations capture ground truth across modalities by identifying \textit{explicit} errors (visual events) and \textit{implicit} errors (tutor cues). Confidence scores allow annotators to flag ambiguity.}
\label{tab:multilabel-annotation}
\fontsize{6pt}{7.6pt}\selectfont
\begin{tabularx}{\textwidth}{| >{\raggedright\arraybackslash}X | c | c | c |}
\hline
\textbf{Multimodal transcription} &
\makecell{\textbf{Student}\\\textbf{error?}\\\textbf{(0/1)}} &
\makecell{\textbf{Implicit/}\\\textbf{explicit}\\\textbf{(0/1)}} &
\makecell{\textbf{Confidence}\\\textbf{level}\\\textbf{(0/1)}} \\
\hline
\textcolor{red}{[14:53]}[\textbf{visual event}]:`Sorry, incorrect...' feedback appears. The correct answer is -12g - 9.1 & 1 & 1 & 1 \\
\hline
\textcolor{red}{[14:55]}[\textbf{visual event}]: New problem loads: -4u + -6u + -2u -5u] & 0 &  & 1 \\
\hline
\textcolor{red}{[14:55]}[\textbf{audio}] \textbf{Tutor}: It's okay. Okay, what about this one? Can we combine or how many terms do we have or are in a relationship with somebody? & 0 &  & 1 \\
\hline
\textcolor{red}{[15:03]}[\textbf{audio}] \textbf{Student}: We have four...actually three, cuz there's three that are negative and one that is positive. & 0 &  & 0 \\
\hline
\textcolor{red}{[15:08]} \textbf{Tutor}: Okay, Does that mean we can combine them or we can't combine them? & 0 &  & 1 \\
\hline
\textcolor{red}{[15:12]} \textbf{Student}: We can. & 0 &  & 1 \\
\hline
\textcolor{red}{[15:13]} \textbf{Tutor}: Yeah, combine them. Perfect. & 0 &  & 1 \\
\hline
\textcolor{red}{[15:15]} \textbf{Tutor}: Okay. So what do we get when we put those all together? & 0 &  & 1 \\
\hline
\textcolor{red}{[15:31]} \textbf{Student}: 17u. & 1 & 0 & 1 \\
\hline
\textcolor{red}{[15:33]}[\textbf{visual event}]: `Sorry, incorrect...' feedback appears. The correct answer is -17u. & 1 & 1 & 0 \\
\hline
\textcolor{red}{[15:36]} \textbf{Tutor}: Not quite. & 1 & 0 & 1 \\
\hline
\end{tabularx}
\end{table}

\vspace{-4mm}

\textit{\textbf{Expert-Based Approaches.}} Rather than using agreement among peer annotators as the benchmark, expert-based approaches treat expert judgment as a more credible reference standard for adjudicating difficult cases \cite{henkel2025humans,ocumpaugh2015baker}. In \cite{nahum2025llms}, researchers implemented an expert-based re-annotation protocol where two experts independently reviewed and reached consensus on difficult cases where models disagreed. This reconciliation process created a high-confidence ``gold standard'' that was demonstrably more reliable than the original labels \cite{na2025llm,wang2024human}. %

\textit{\textbf{Predictive Validity.}} Predictive validity ties the value of an annotation to its ability to forecast performance on an external, related measure. This concept is rooted in process-product research \cite{rosenshine1970evaluation} and foundational AIED frameworks like Knowledge Component Analysis, where a model is validated by its ability to predict student error rates \cite{koedinger2012knowledge}. For example, an AI score on an open response gains validity if it accurately predicts the score on a corresponding MCQ \cite{thomas2025does,thomas2025improving}. However, predictive validity must be interpreted with caution. A high correlation with an outcome does not guarantee that the correct construct has been identified, e.g., incorrectly labeling ``frustration'' as ``confusion'' might still yield a predictive correlation with low performance because the two states are often collinear. Predictive validity should not be used in isolation but as a method to triangulate construct validity \cite{self1990theoretical}. Researchers should verify that specific labels predict distinct outcomes (e.g., does the label predict the \textit{specific} type of error?) to ensure the model is capturing the intended construct, not just a signal \cite{Rachatasumrit2024Beyond}.

\textit{\textbf{Close-the-Loop (Intervention-Level) Evidence.}} Close-the-loop evidence assesses whether a given annotation scheme, and the AI system trained on it, produces improved student learning. For instance, Tutor CoPilot classifies pedagogical moves to give advice to tutors. It was evaluated by comparing student learning outcomes when tutors receive AI-generated advice versus control conditions. This ``closes the loop'' between the annotation, the AI system, and the desired educational outcome \cite{wang2024tutor}. While this ``closes the loop'' between annotation and outcome, strong construct validity is still needed to ensure the learning gains are attributable to the specific pedagogical constructs the AI identified, not confounding factors \cite{trochim2016research}. By integrating these validity-centered frameworks, in addition to including traditional reliability estimates, researchers can move beyond simply measuring agreement. This shift prioritizes the goal of AIED: creating tools that are genuinely effective, not just consistent.

\section{Next Steps for AIED \& Conclusion}

To move from the theoretical shifts proposed in this paper to a new standard of practice, the AIED community should consider: 1) \textbf{``uncertainty-aware'' infrastructure} and 2) \textbf{an evolution in peer-review culture.}  First, while privacy constraints often limit the sharing of raw data, the community could benefit from adopting standardized reporting frameworks that capture ``productive ambiguity.'' Researchers should consider documenting de-identified edge cases and the specific rationales behind expert disagreements to illustrate construct complexity. As demonstrated in Table 4, integrating annotator confidence levels and multi-label schemas allows for a more nuanced understanding of ``ground truth'' that acknowledges the inherent subjectivity of high-inference tutoring interactions. Second, peer-review could gradually move toward prioritizing the transparent reporting of rater expertise and predictive validity evidence, such as demonstrating that the labels used to train a system effectively forecast learning outcomes. Rather than treating reliability as a mechanical gatekeeper (e.g., $\kappa$ > 0.8), reviewers are encouraged to treat IRR as a diagnostic signal used to localize disagreement and refine pedagogical constructs (Table 3). This shift ensures that labels are validated not just by statistical agreement, but by their demonstrated alignment with meaningful instructional mechanisms and learning outcomes.

The rapid adoption of GenAI has increased the stakes of how we define and evaluate the quality of labeled data. By shifting the focus from mechanical consensus to diagnostic rigor, the AIED community can ensure that models are not merely mimicking human consistency, but are accurately capturing the complex, ``noisy,'' and meaningful processes that drive learning. Our goal is not to replace established psychometric frameworks, but to translate and extend them for a new era of AI-driven education, characterized by high-inference constructs and multimodal data often annotated by LLMs.

\begin{credits}
\subsubsection{\ackname}
This work was supported in part by the Learning Engineering Virtual Institute, the Gates Foundation, and the Chan Zuckerberg Initiative. The opinions, findings, and conclusions expressed are those of the authors.
\end{credits}

\bibliographystyle{splncs04}
\bibliography{main} %

\end{document}